\documentclass[10pt,journal,compsoc]{IEEEtran}

\usepackage{amssymb}
\usepackage{subfig}
\usepackage{float}
\usepackage{balance}
\usepackage{graphicx}
\usepackage{color}
\usepackage{url}
\usepackage{times}
\usepackage{amsmath}

\usepackage[font={footnotesize}]{caption}

\newcommand{\trainInd}{}
\newcommand{\myparagraph}[1]{\vspace{4pt} \noindent \textbf{#1.}}

\ifCLASSOPTIONcompsoc
  \usepackage[nocompress]{cite}
\else
  \usepackage{cite}
\fi

\begin{document}
\title{Using EEG-Based BCI Devices to Subliminally Probe for Private Information}

\author{
Mario~Frank, Tiffany~Hwu, Sakshi~Jain, Robert~T.~Knight, Ivan~Martinovic,\\ Prateek~Mittal, Daniele~Perito, Ivo~Sluganovic and Dawn~Song}

\IEEEtitleabstractindextext{

\begin{abstract}
EEG-based Brain-Computer-Interfaces are becoming available as consumer-grade devices, used in applications from gaming to learning programs with neuro-feedback loops.
While enabling attractive applications, their proliferation introduces novel privacy concerns and security threats. One such example are attacks in which adversaries compromise EEG-based BCI devices, and are able to analyze the user’s brain activity to infer private information about a user, such as their bank or area-of-living. However, a
key limitation of the above attacks is that they require user cooperation, and are thus easily detectable and rendered inefﬁcient after
discovery.

In this paper, we propose and analyze a more serious threat - a subliminal attack in which, given that the visual probing lasts for less
than 13.3 milliseconds, the existence of any stimulus is below ones cognitive perception. We show that, even under such strong
limitations, the attackers can still analyze subliminal brain activity in response to the rapid visual stimuli and consequently infer private
information about the user.

By running a proof-of-concept study with 27 participants, we experimentally evaluate the feasibility of subliminal attacks using
EEG-based BCI devices. While not perfect, our results show that it is indeed feasible for attackers to subliminally learn probabilistic
information about their victims.

\end{abstract}

}

\maketitle

\IEEEdisplaynontitleabstractindextext

\IEEEpeerreviewmaketitle

\section{Introduction}

Brain-Computer Interface (BCI) devices are becoming increasingly popular for use in applications such as entertainment, accessibility, and cognitive enhancement~\cite{neuroptimal,smartbrain}.
A popular technology used in BCI for recording brain activity is Electroencephalography (EEG), which uses external scalp electrodes to capture fluctuations of the electrical potentials in the brain.
The recorded EEG signal is processed by the supporting software, which extracts salient brainwave features, translates them into specific computer instructions, and feeds them back to the invoking application.
The Emotiv device~\cite{emotiv} and the Neurosky device~\cite{neurosky} are examples of low-cost commodity BCIs, intended for home usage with applications written by third-party developers and are available for download from application markets (see, e.g.,~\cite{store-emotiv, store-neurosky}).

\myparagraph{Privacy leaks via BCI}
Martinovic et al.~\cite{Usenix12_EEGattacks} recently emphasized that BCI devices may make the raw EEG signal available to potentially untrusted third-party applications.
If such an application is malicious, it can in turn abuse the BCI device to infer private information about a victim, such as her/his preferred bank or area-of-living.
The general idea of this attack is similar to a polygraph, where the interrogated person's physiological reactions are used to reason about his/her knowledge.

\myparagraph{Limitations of supraliminal stimuli}
However, a fundamental limitation of the attacks proposed by Martinovic et al. is that they rely on supraliminal (consciously perceived) stimuli and are thus detectable.
In the proposed attack, victims are repeatedly exposed to specific visual stimuli that elicit distinctive and measurable cognitive responses, resulting in leakage of private information, such as user's month of birth or chosen bank.
Given the duration and repetitiveness of visual stimulation (sequence of images shown over 90 seconds), users would easily become suspicious and detect all but the most successfully integrated and innocuous probes.
Another issue preventing wide deployment of this attack is that after a few users detect abnormal behavior of their newly downloaded application, they would report it and flag it as not functional or even malicious.
It is a reasonable assumption that all users of the application are connected via the application market, which enables sharing such warnings or flags.
This would prevent the attacker from carrying out a large scale attack.

\vspace{6pt} \noindent \textbf{Can user privacy be compromised subliminally?}
Based on these observations, we focus this work on researching if the threat and potential reach of previously proposed attacks can be significantly increased by preventing their conscious detection.
As a result, we propose a \emph{subliminal} attack that infers private information by probing the victim at a level below his/her cognitive perception.
Similar to \emph{subliminal advertising} (see, e.g.,~\cite{subliminal-advert}), our key idea is to show the visual stimuli within the screen content that the user expects to see, but for a duration that is too short for conscious perception (several milliseconds), yet still sufficient to result in detectable activation of certain parts of user's brain.
As an example, for a video-based application, we propose to implement small snippets of visual stimuli within a few frames of the video.
The attacker succeeds if he can infer private information about the user without arousing the user's suspicion.

This is a challenging task.
If the stimuli are shown too prominently, this increases the chance of the attack being detected.
If, in contrast, the attacker does too good a job of hiding the stimuli, the user's subliminal detection may not be sufficiently strong, reducing the probability of inferring relevant private information.
Thus, the attacker must operate within this narrow regime of the user's input channel.

\myparagraph{Results}
In this paper, we experimentally demonstrate for the first time the feasibility of a subliminal attack via a proof-of-concept study with 27 subjects.
We conduct experiments on users wearing EEG-based BCI devices.
Our study implements an attack scenario where a user watches a video and the attacker tries to infer whether the user recognizes a particular person by hiding pictures of this person in the video for a time duration shorter than 13.3 milliseconds.
To analyze users' subliminal reactions to the embedded visual stimuli, we use machine learning techniques on the recorded EEG signal.
For 18 out of the 27 subjects, our classifier was able to guess the secret correctly, thus \emph{reducing the attacker's guessing entropy by 20.8\%}.
This ratio even increases to 8 out of 9 in one variant of the attack, which \emph{reduces the attacker's entropy by as much as 48.8\%} when compared to random guessing.
The success chance did not vary significantly between subjects who were able to detect that something was hidden in the video and subjects who did not notice anything abnormal.
While not perfect, these experimental results show that the subliminal attack is indeed feasible -- attackers can make probabilistic inferences on the users' recognition of the person depicted in the visual stimuli in a manner that is concealed from the user.

\noindent
In conclusion, this paper makes the following \textbf{contributions:} 
\begin{itemize}
\item We propose a new attack against users wearing BCI devices that is subliminal: it infers users' private information by exploiting brain activity in response to visual stimuli that are not cognitively perceived by users.
    When carried out carefully,
    such an attack can remain undetected.

  \item We experimentally demonstrate for the first time the feasibility of subliminal attacks on EEG-based BCI devices via a user study of 27 subjects.
    Our experimental results show the feasibility of subliminally learning private information about a user, such as whether the user recognizes the person depicted in the visual stimuli.

  \item We discuss several potential countermeasures against the presented attacks which subliminally probe for BCI users' private information.

\end{itemize}

\begin{figure}[t]
  \centering
  \includegraphics[trim={0 8.5cm 0 0}, width=0.8\columnwidth]{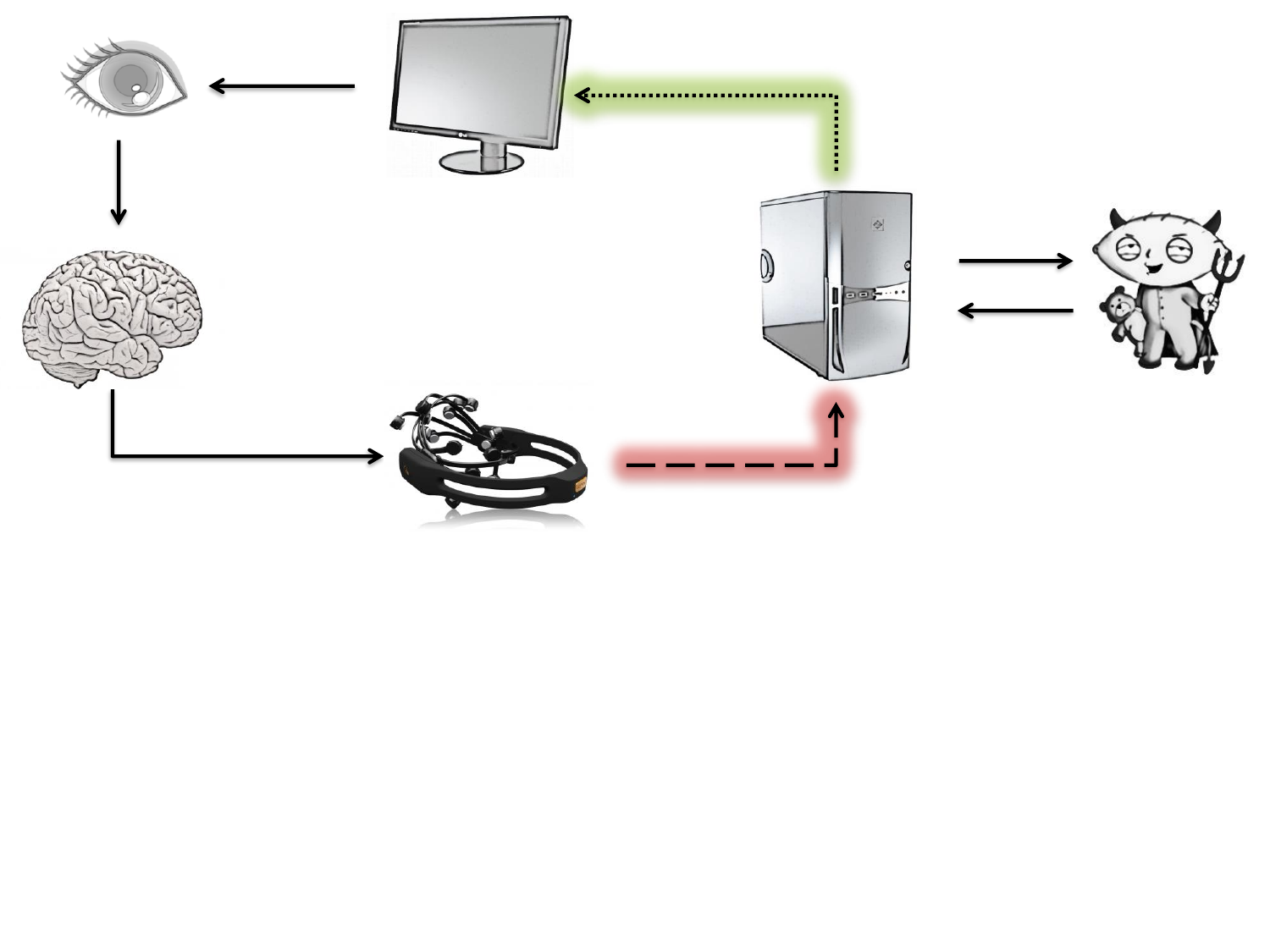}
  \caption{
    The communication channels between  the computer, the user, and the BCI headset.
    The investigated side-channel attack is based on malware that records data from the channel between the BCI headset and the computer (dashed) and analyses it in order to uncover user's private information.
  This attack also contributes a signal to the visual channel (dotted), however this is done in a way that prevents the user to consciously notice the modifications.
  } \label{fig_HCIchannels}
\end{figure}

\section{Neuroscientific Background} \label{sec:neuroBackground}
\myparagraph{EEG-based BCI}
Electroencephalography (EEG) monitors electrical activity at the scalp that corresponds to changes in ion concentrations of neurons in a functioning brain. A typical use of EEG involves the attachment of one or several electrodes to coded locations of the scalp and monitoring changes in potentials. The signal of each pair of electrodes is amplified through a differential amplifier, filtered, recorded at a high sample rate (typically in the range of 128Hz-16kHz), and saved for later analysis.

EEG is widely used in a medical setting to monitor neurological diseases.
For instance, patients with epilepsy often undergo EEG to observe and categorize seizures, which aids in the appropriate choice of treatment.
Other medical uses include diagnosing possible brain death of comatose patients, communication with patients who suffer from locked-in syndrome, or control of wheelchairs for handicapped~\cite{Carlson2013}.

In neuroscience research, EEG serves as a non-invasive, cost-effective method of measuring brain activity.
In comparison to other methods, such as functional MRI, EEG is relatively low in spatial resolution, but very high in temporal resolution, able to precisely capture millisecond-scale changes in the activity of a specific brain regions.

Depending on the needs of the users, EEG recording devices vary in sampling frequency, number and location of electrodes, and overall signal quality.
Recently, affordable and portable EEG devices have appeared on the market in the form of lightweight consumer products for gaming and personal EEG monitoring.
Cognitive states picked up by such devices can be utilized by games to allow users to control on-screen avatars, monitor and train their own mental states, and improve gaming experience by collecting information on reactions and emotions.

\myparagraph{Event-Related Potentials (ERP)}
In many applications, EEG signal is analyzed in conjunction with the presentation of visual or auditory stimuli.
The consequent EEG waveform caused by the presentation of stimuli can be categorized into event-related potentials (ERP), which are combinations of negative and positive spikes occurring at different times after certain types of stimuli.
Those ERP-s that reliably occur after stimuli onset are usually named by the polarity of the wave (N or P) and their delay in milliseconds.

In our research, the most relevant example of an ERP is the P300, a positive amplitude response whose amplitude peak occurs approximately 300~milliseconds after stimuli onset.
While both its amplitude and latency can vary significantly (from as much to 250~ms to 500~ms), P300 was shown to be consistently elicited when individuals decide between relevant and irrelevant stimuli, including images or sounds considered novel or threatening~\cite{Fogelson:2009:MEL:1618368.1618372,Knight96}.
The P300 has been successfully used in EEG devices to enable users to spell letters using only EEG signals~\cite{Hoffmann} and to successfully perform \emph{guilty-knowledge} tests, despite user's active efforts to conceal~\cite{Farwell2001}.

While in our attack we do focus on a single specific ERP, given the relevance of P300 to the attacker's goals of extracting relevant information and previously achieved successes in its use as a guilty-knowledge test, in our analysis we use it as a guidance to choose a small subset of EEG channels used by the classifier.

\myparagraph{Challenges of subliminal stimulation}
While the majority of sensory stimulation that we are exposed to in everyday lives is sufficiently intensive to be consciously perceived (\textbf{supraliminal}), in certain situations a stimulus can also be made \textbf{subliminal} if its intensity is carefully controlled.
For instance, if a visual stimuli is shown sufficiently briefly, individuals might not be consciously aware of it, but research has shown that such subliminal stimuli does elicit responses as the brain processes it.
Furthermore, it was shown that subliminal stimuli even measurably impact one's behavior, for instance by influencing the choice of consumer brands~\cite{Karremans2006}.
Given that an individual might not even be aware of the influence that they are exposed to, the use of subliminal stimuli in advertising purposes is strictly regulated in many countries.

\begin{figure}[htb]
	\centering
	\includegraphics[width=0.30\textwidth]{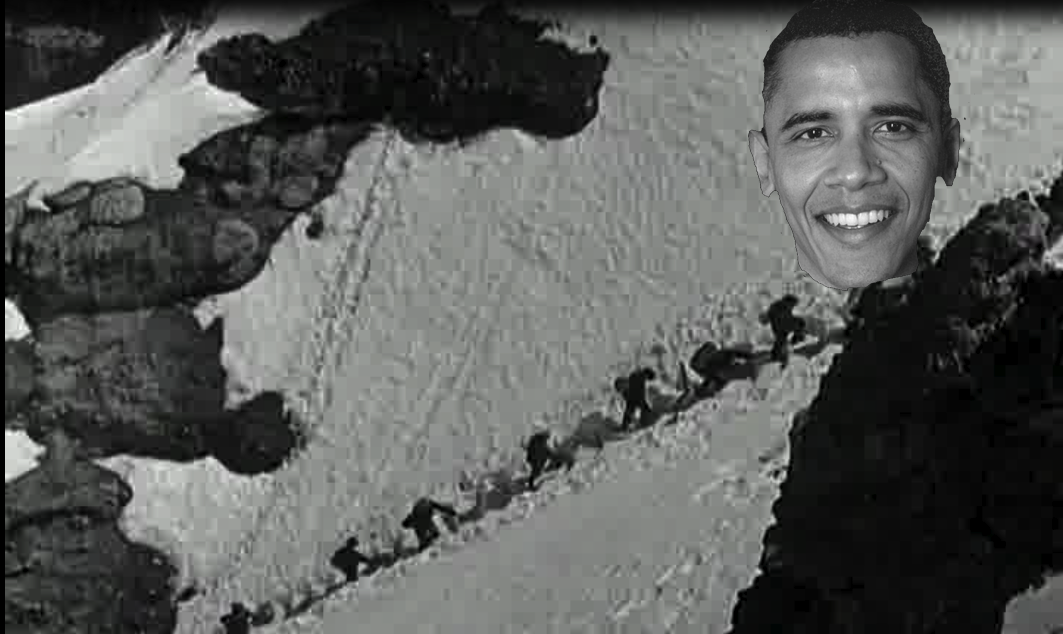}
	\caption{
		An example of a video frame with the hidden subliminal stimulus in the form of a face whose recognition the attacker wants to probe.
		Shown only for the duration of a single frame, the stimulus is below cognitive perception for most individuals.
	}
	\label{fig_videos}
\end{figure}

Researchers agree on the existence of \emph{perceptual threshold}, an intensity that defines whether one will cognitively perceived a stimulation or not, but determining specific values for any given stimulus type is not straightforward.
For instance, neuroscience literature speaks of 10~ms to 55~ms as a suggested presentation time range for a stimulus to be subliminal (a good overview of designing experiments with subliminal stimuli can be found in~\cite{subliminal-experiments}), but it is also known that this duration not only varies significantly among individuals, but also changes from day to day for the same individual.
As a result, most definitions of perceptual threshold focus on levels of stimulation that result in stimulus being undetected in a certain percentage of times it was presented to a user.

In order to allow comparisons, we must necessarily use a single stimulus duration for all participants.
Considering the limitations of the refresh frequencies of common off-the-shelf displays, we chose to show each subliminal image for 13.3~milliseconds.
While this is on the shorter side of commonly used timings for subliminal visual stimulation, we believe that it will enable us to show the stimulus multiple times and still remain subliminal to some users.
On the other hand, shorter stimulus is likely to result in weaker brain responses, which makes our task of extracting private information additionally harder, and as such makes any positive results significant.

Finally, we emphasize that an attacker could adapt the stimulus duration to each specific victim and thus both achieve better hiding of the stimuli, as well as stronger brain responses.
We further reflect on this in our results analysis in Section~\ref{sec_discuss}.

\section{Threat Models and Attacks} \label{sec_attacker}
In this section we introduce the system and adversary models and describe the attack scenario for subliminal probing of private information.

The system model consists of the user, the computer, and the BCI-device.
The user has bought a device and uses it at home with various applications downloaded from the third-party developer platform.
The user trusts the computer and the BCI device.
The user actively supports setting up the device and calibrates it, if necessary in order to ensure the minimal signal noise.
We assume that the user has an incentive to use the downloaded applications.

The attacker is an application developer.
Through BCI-device's API, the application can access the raw EEG signal recorded by the device.
The attacker has modified a benign game or video viewer by inserting his malicious code and has posted the application on the platform with a slightly modified application name.
We assume that executing attacker's application does not result in the overall compromise of the system, i.e. attacker is not able to break out of the BCI application's sandbox.

The goal of the attacker is to obtain private user information.
This could be any of the scenarios proposed in~\cite{Usenix12_EEGattacks} such as guessing the banking provider, PINs, or month of birth.
Generally, the target can be any memory of the user that could be useful for an attacker.
For instance, the attacker can subliminally probe (until successful) and blackmail users who seem to be familiar with the logos of particular porn sites, or a repressive regime can try to identify which users are familiar with some of the key persons of the underground opposition.

\myparagraph{Attack preparation}
The attacker uses event-related potentials (ERPs) to run the following strategy.
The attacker designs a number of visual stimuli that correspond to alternative answers of the questions to which he wants to find an answer.
For instance, the attacker wants to verify if the user recognizes a particular person, so he places this image and other visual stimuli, such as images of other people, at random times and random positions of the application and offers the application in the online application store.
In our setting, the application is a video viewer, and the attacker includes images of people at random frames and at random screen locations of the video.
See, for instance, the video frame depicted in Figure~\ref{fig_videos}.
In this paper, we assume that the attacker wants to find out if the suspect recognizes a specific the face.

\myparagraph{Attack execution}
After the user has downloaded and installed the application and starts using it, the attacker collects the EEG signal recorded while the user is exposed to the different images displayed on the screen.
Hoping that the person known to the user triggers the strongest event-related potentials, the attacker analyzes the EEG signal in a comparative way.
This analysis works best if the recorded data can be contrasted with prior recordings of the victim, where it is known what the most relevant stimulus was.
An example of suitable prior observations for this purpose is EEG data that has been generated while the user was calibrating the BCI device.
The usual calibration step consists of a sequence of numbers that are being flashed randomly. The user must count the occurrence of a particular number.
Waiting for this number to appear is a very good way to provoke P300 artifacts.
This calibration step of the device is often used by benign applications that rely on ERPs, too.
As a result, the attacker can incorporate his attack in an application that requires this calibration step anyway.

\begin{figure}[tb]
	\centering
	\includegraphics[trim={0cm 11cm 0cm 0cm}, width=1\columnwidth]{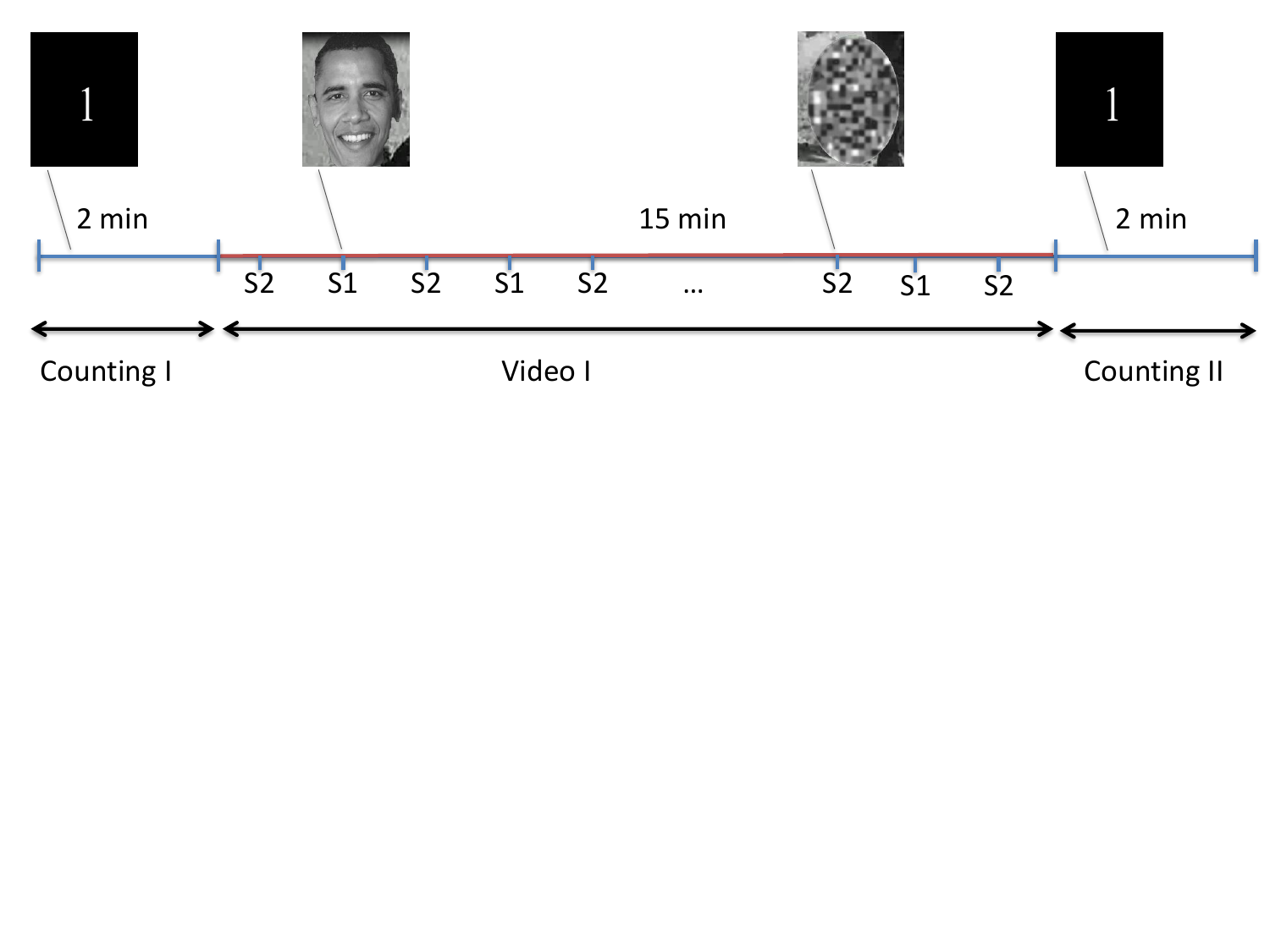}
	\caption{
		The experimental protocol is divided into 3 sub-experiments: Counting I, Video I and Counting II.
		The embedded visual stimuli $S_j$ are depicted above the timeline.
	}
	\label{fig_ExpProtocol}
\end{figure}

\section{Experiments}  \label{sec_exps}
In this section we describe the setup of our proof-of-concept experiment.
Investigating whether the proposed subconscious side-channel attack is feasible is a very challenging task due to many factors that can affect the results.
A negative outcome could have many reasons.
For instance, the equipment used could be suboptimal, the video used to hide the attack could have many still
images where it is hard to hide a stimulus, the secret that is being attacked could be too complex, and so on.
Therefore, our experiment is designed to investigate whether the attack is feasible instead of starting off with sophisticated variants.
In particular, we make design decisions that minimize the chance that the attack fails due to factors that we can control.
For instance, we use a good EEG device and a video with flickering artifacts that helps hide the attack.
The following sections detail other design decisions.

\subsection{Test Population and Setup}
After obtaining approval from the Institutional Review Board, 29 
undergraduate and graduate students (21 males and 8 females)  in the 
Computer Science department were recruited to participate in our experiment, 2 of which had unusable data due to recording problems.
All subjects were self-screened for neurological disorders and metal implants which could potentially interfere with recording.
Prior to the experiment, subjects were informed of the basic EEG procedures, but not yet informed of the subliminal nature of the stimuli.
The participants signed informed consent and received compensation in the form of a \$40 gift card.
The experiment took 90 minutes total for each user, including setup time.
This  was the main limiting factor for population size.
ActiveTwo BioSemi equipment~\cite{biosemi} was used for the collection of EEG data.
Participants were measured and fitted with a tight cap, and 64 Ag/AgCl electrodes were attached to the cap with conducting gel. 
All electrodes were then attached to a low-noise DC coupled post-amplifier, with a sampling rate of 1024 Hz.  All stimuli were presented in a dim room on a CRT monitor (75Hz refresh rate) using presentation software~\cite{presentation}.

\subsection{Experimental Protocol}
After the setup described above, the participants were asked to try to remain relaxed
for the entire duration of the experiments. 
 The interaction with the participants was kept as short and concise as possible. There were four
parts to the complete experiment, two parts pertained to a counting task discussed below, and the other
task involved observing videos.
Figure~\ref{fig_ExpProtocol} semantically shows the parts on a timeline for better understanding.
The following sections give details of each part of the experiment.

\myparagraph{Counting I and II}
In these parts of the experiment, the participant was presented with a randomly permuted sequence of numbers from 0 to 10. Each number except 1 appeared exactly 16 times. The digit 1
could appear anywhere between 14-18 times, chosen uniformly at random. The participant was asked to count the number of occurrences of the number 1. Each stimulus lasted for 250~ms, and pauses between stimuli were randomly chosen to be between 250~ms
and 375~ms long. At the end of this step of the experiment, the participants were asked for their count to check for correctness.
This part of the experiment lasted for about 2 minutes. It was carried out in the beginning of the experiment (counting I) and at the end (counting II). These counting tasks are standard
tasks used to calibrate BCI devices to users to ensure correct functionality of BCI applications.

\myparagraph{Video I}
In this phase, the participant was instructed to watch a 15 minute long black and white video extracted from Charlie Chaplin's "The Gold Rush" (1925).
They were asked to pay attention to the plot of the video to make sure they concentrated on watching the video through its entire duration.
Two kinds of stimuli (S1 and S2) were used, one with a black and white portrait of Barack Obama (Figure~\ref{fig_ExpProtocol}) (S1) and the other being a blurred image of a human face (S2).
We choose these stimuli in order to make sure that every subject was familiar with S1 and would not recognize S2.
Given that the our experiments were conducted at a US university at the time when Barack Obama was the acting president, we can safely assume that we know the correct answer for all participants. 
In a real-world scenario, the adversary might similarly prove victim's recognition of a known face in order to validate a particular configuration of the attack.

A stimulus was shown every 5 seconds, making a total of 180 stimuli over 15 minutes.
Every 4$^{th}$ stimulus was S1 and was displayed at the top right corner of the image frame.
The position of S2 rotated along the remaining three corners.
Each stimulus was shown for 13.3~ms.
The limiting factor of this time was the screen refresh rate (75Hz).
Once the video ended, the participant moved on to the next part of the experiment.

\myparagraph{Recognition survey}
As we are ultimately interested in understanding the feasibility of carrying out the attack subconsciously, the participants were asked at the end of the experiment if they noticed anything odd in the video.
If they negated, no further questions were asked.
However, in the case of a positive answer, they were asked for details of what they saw.
We categorize their answers as follows: participant recognized nothing, participant saw something, participant saw a face, participant saw ``Barack Obama".

\myparagraph{User study limitations}
Considering that we designed the user study to mimic a potential real-world attack, while at the same time minimizing the likelihood of a negative result due to controllable factors, it does have a few limitations that we discuss upfront.

Firstly, the study focuses on probing a single type of private information: whether the user recognizes a given face or not.
While this does not allow us to reason about the ability to extract arbitrary information, it does indicate that similar subliminal attacks might be possible.
Secondly, the subliminal nature of our study does not allow us to fully disclose its details to participants until its very end.
As a result, our only measure of the extent to which a particular user detected any probing is the recognition survey.
While we made several steps to reduce the experimenter bias and pose the question of "noticing anything odd" neutrally, it is possible that the responses of some participants were influenced by the mere fact that they were being asked such a question, or by wanting to please the experimenters by giving a positive answer.
However, we believe that participants are less likely to mis-report not seeing anything than to conform to potentially detected demand characteristics~\cite{orne2009demand}.
Additionally, the survey answer was not properly noted for four participants; those were indicated as N/A in our results.
Lastly, even though we screen for neurological disorders, some conditions, such as Developmental Prosopagnosia~\cite{duchaine2005dissociations}, inhibit face recognition, but often remain undetected.
Despite the study being carried out at a US university during Barack Obama's presidency, it is possible that some users were unable to recognize his face and thus impact the final results.

Despite these limitations, the results of our user study provide a clear confirmation that subliminal probing for private information is indeed possible, and could present a significant threat to user privacy in the future.

\section{Data analysis} \label{sec_meth}

The methods described in this section serve to investigate the level to which different subliminal side-channel attacks with BCIs are feasible if the attacker controls screen content and has access to raw EEG data.
We are carrying out a proof-of-concept by running an attack under controlled conditions.

As described above, the user has calibrated the BCI by actively participating in a counting experiment.
Then, at a later point in time, the user watches a video while still wearing the device.
The attacker manipulates the video and insert oval images of two kinds at random screen locations and at random times.
His goal is to identify which person does the user recognize by analyzing the EEG data that is collected while the user watches the video.
From a technical perspective, we want to evaluate whether the classifier can extract sufficient information from the recorded EEG signal in order to determine one of the three different types of brain activity: 1) unknown face, 2) a face that the user subliminally recognizes, or 3) a plain video sequence without any subliminal stimulation.
Given a classifier that consistently discriminates between these three types of brain activity, the attacker can use it to learn new private information about a chosen victim.

\subsection{Data Acquisition and Preprocessing}
\begin{figure}[tb]
	\centering
	\includegraphics[width=0.85\columnwidth, trim={2cm 1.3cm 0 0.3cm}]
	{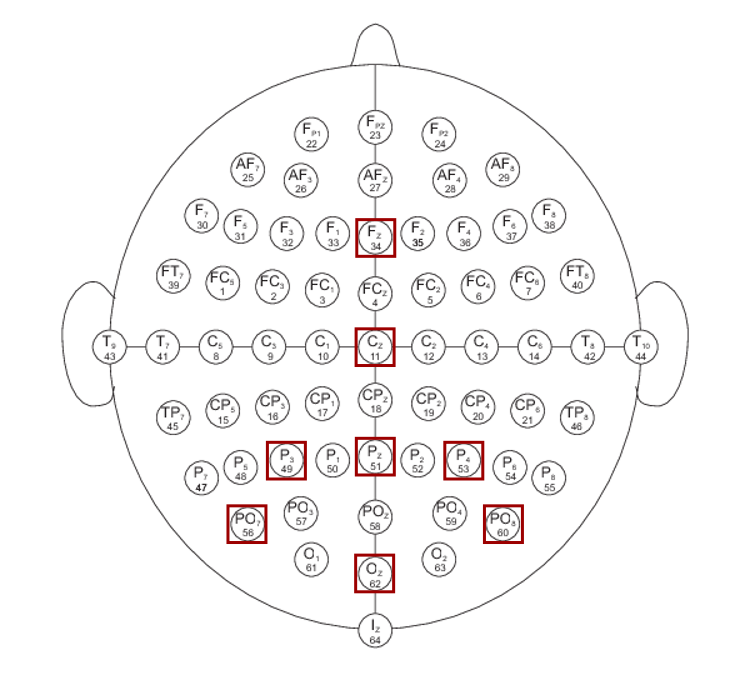}
	\caption{
		Positions of the channels used by the classifier.
		The classifier uses primarily data from parietal and occipital lobes.
	}
	\label{fig:channelPositions}
\end{figure}

The raw data consists of wave signals from a number of different electrodes, called channels, whose positions are shown in Figure~\ref{fig:channelPositions}.
The subset of channels used by the classifier is marked with red rectangles, and each channel is sampled at 1024Hz.
We correlate the EEG signals with exact positions of stimuli using the timestamps obtained from the software.
For preprocessing, we first divide the signal into epochs, each epoch ranging from 200~ms prior to 1000~ms after every stimulus onset.
Each such epoch is associated with the respective stimulus that triggers it.
For each epoch, we then calculate the mean of the first 200~ms to get a baseline and subtract this baseline from the entire epoch.
We reduce the high frequency noise by passing the signals through a low pass filter with a pass band of [0.35, 0.4] in normalized frequency units.
Finally, we apply a median filter that extracts the median from each four consecutive measurements.

\subsection{Classification} \label{sec_class}
The goal of the attacker is to train a classifier which identifies if the stimulus is relevant for the user.
Classification requires a training set of data containing observations whose class membership is already known and a test set to evaluate its performance.
We now describe the classifier framework that we use to predict stimulus relevance from EEG signals.

\myparagraph{Classification Setting}
In our setting, each epoch is one observation.
Each epoch corresponds to a single stimulus and contains the  signals from all the EEG channels for a time period of [signal - 200ms, signal + 1000~ms]. If $C$ denotes the number of channels being used and $f$ denotes the sampling frequency, we have $(1000 + 200)f = S$ measurements per channel per epoch.
We group the signals from all the channels for each epoch into a  feature vector of dimensionality $K = C \times S$.
The classification algorithm consists of two phases, training and testing.
In the training phase, the input samples provided to the classifier are of the form: $\{x_i\trainInd \in X , y_i\}$ where $X = \{x_i \in \mathbb{R}^K, i = 1,...N\}$, N denotes the total number of input epochs. The set $Y = \{y_i \in \{0,1\}, i = 1,...N\}$ denotes the class labels, i.e. whether epoch $x_i\trainInd$ corresponds to the target stimulus relevant to the subject ($y_i = 1$) or not ($y_i = 0$). Since the system used to display the stimuli captured the indicator of each stimuli and the corresponding timestamp, for each epoch, the value of $y_i$ could be obtained. Given this set of inputs for training, the classifier learns the function that maps the feature vector (epochs) to the stimulus indicator:
\begin{equation*}
f(x_i): \mathbf{x_i} \in \mathbb{R}^K \rightarrow \mathbf{y} \in \{0,1\}
\end{equation*}

In the testing phase, the classifier is provided with a set of fresh observations $\mathbf{x_i}$ for which it must output label predictions $\mathbf{y_i}$. In other words, the classifier must predict for each epoch, if the corresponding stimulus shown to the participant is relevant or not.

\myparagraph{Boosted Logistic Regression (BLR)}
We use a logistic regression method to make predictions on the stimulus relevance given the EEG signal. We train this classifier by minimizing the negative Bernoulli log-likelihood of the corresponding model in an iterative fashion as proposed in~\cite{Friedman98additivelogistic, Friedman00greedyfunction}.  A variant of this technique was used for P300 spelling in~\cite{Hoffmann} with MATLAB code being available online. Also, Martinovic et al. used it for their attack~\cite{Usenix12_EEGattacks} and it showed good performance for guessing user secrets from event-related potentials. For these reasons, we also chose this classifier for our experiment. As follows, we briefly describe how the classifier works. We will use the same notation as in the original paper~\cite{Hoffmann} to simplify following up on details.

The boosted logistic regression classifier (BLR) variant that we employ is an ensemble method. This means it consists out of a set of $M\in\mathbb N$ individual classifiers $f_m$ with $m\in\{1, \dots ,M \}$ that all output individual classifier scores.
Each classifier has a linear form $ f_m(\mathbf{x}_i\trainInd; \mathbf{w}_m) = \mathbf{w}_m^T\mathbf{x}_i\trainInd$ with coefficients $\mathbf{w}$ that differently weight  the recorded channels at different points of time.
These individual classifiers are incrementally blended into a single classifier $F_m$.
At step $m$, the probabilistic model underlying classifier $F_m$ is
 \begin{equation}
 p_m(y_i\!=\!1\vert \mathbf{x}_i\trainInd) = \frac{\exp(F_m(\mathbf{x}_i\trainInd))}{\exp(F_m(\mathbf{x}_i\trainInd)) + \exp(-F_m(\mathbf{x}_i\trainInd))}
 \end{equation}
and represents the probability that the stimulus of the current epoch $i$ is a target stimulus given the concatenated EEG signal $\mathbf{x}_i\trainInd$.
The likelihood of this model for all epochs of the training data (i.e. the data where labels $Y_i$ are given) is
\begin{equation}
  \begin{split}
    L(F_m; \mathbf{X\trainInd}\!, \mathbf{Y}) \!=\! \prod_{i=1}^{N}
    p_m(y_i\!=\!1\vert \mathbf{x}_i\trainInd)^{y_i} \cdot [1 \!-\!   p_m(y_i\!=\!1\vert \mathbf{x}_i\trainInd)]^{1-y_i}
  \label{eq_boostLogReg}
  \end{split}
\end{equation}

In the training phase, we iteratively optimize this function.
At every optimization step $m$, the classifier of the last step is updated by adding a new weak classifier: $F_m=F_{m-1}+ f_m$.
The weights of this additional classifier are computed by minimizing the least-squares distance of the gradient of the log-likelihood:
\begin{equation}
f_m \!\!= \!\!\arg\!\min_{f} \sum_{i=1}^{N}  \!\!\left( \!\!\left[ \frac{\partial L(F(\mathbf{x}_i\trainInd))}{\partial F(\mathbf{x}_i\trainInd)} \right]_{\!\!F=F_{m-1}}
 \!\!\!\!\!\!\!\!- \!f_m\!(\mathbf{x}_i\trainInd; \mathbf{w}_m)\!\! \right)^{\!\!2}
\label{eq_weakLcostFunc}
\end{equation}
At each update step, the new weak classifier $f_m$ is added with a weight $  \gamma_m$ such that, finally, the ensemble classifier is $F_M= \sum_{m=1}^M  \gamma_m f_m$. These weights are computed  after the optimization step of the respective  $f_m$ with Eq.~\ref{eq_weakLcostFunc}. It is selected  such that  Eq.~(\ref{eq_boostLogReg}) becomes maximal.
Please see the full details of this algorithm and an experimental evaluation in ~\cite{Hoffmann}.

\myparagraph{Data Dimensionality \& Features}
In the described training phase, we provide all EEG data of the counting experiment together with the class labels to Eq.~(\ref{eq_boostLogReg}).
For instance, if the classifier is trained to detect recognition of number 1, then epochs triggered by the number 1 are accounted to class $y_i=1$ and epochs with other numbers get the class label $y_i=0$.
For each weak classifier $C\cdot S$, we must learn the coefficients $\mathbf{w} \in \mathbb R^{K}$.
If we record many channels at a high frame rate, then the optimization in Eq.~(\ref{eq_weakLcostFunc}) can become under-determined if too few epochs are available for training.
We approached this problem of a low observation-to-dimension ratio by taking only those channels into account that are located along the z-axis, parietal, and occipital areas of the scalp where P300 ERPs are usually the strongest. In particular we use only channels `Fz', `Cz', `Pz', `P3', `P4', `PO7', `PO8', `Oz'.
In order to further reduce the dimensionality of the data, for each channel we compute the median of every 4 consecutive measurements to reduce the dimensionality by a factor of 4. This adds the side effect of denoising  the signal by a median filter.

\section{Evaluation} \label{sec_results}
In this section, we evaluate the feasibility of subliminally probing for private user information by running several experiments, each representing a different scenario that an attacker might attempt.

We first confirm that the described classifier indeed achieves results that are comparable to previous work on EEG-based attacks when trained and tested on supraliminal data.
We then show that our subliminal stimuli has indeed remained hidden for a subset of users.
As a result, subliminal probing of some users is possible if their brain responses are sufficiently strong.
This threat is confirmed by showing in Section~\ref{ssec:probingSubliminalInfo} that a classifier trained on subliminal data can achieve strong classification results when evaluated on the same type of data.

Most importantly, Section~\ref{ssec:transferLearning} shows that an attacker can train the classifier on data recorded during supraliminal stimulation, such as the one used during the calibration of the BCI device, and then use it to probabilistically learn private information about the victim who is subliminally probed.
Finally, we describe and show how can an attacker additionally improve the classification performance, given that he has an estimate of the classifier's confidence.

\subsection{Classifier Validation on Supraliminal Stimuli} \label{ssec:classifierValidation}
The success of the attack presented in Section~\ref{sec_attacker} depends fundamentally on the strength of the BLR classifier described in Section~\ref{sec_class}.
Before evaluating it on more difficult subliminal attacks, we begin by affirming its reliability using a sanity check by both training and testing on supraliminal data.

\myparagraph{Setup}
In this experiment, we train BLR with data collected from the first counting phase of the experiment.
This is the phase where the user calibrates the device by counting the occurrences of number 1, while a sequence of random numbers between 0 and 10 is repeatedly flashed on the screen.
We extract the corresponding epochs from the EEG recordings and label all epochs that correspond to number 1 as the positive class, while all other epochs are labeled as the negative class.
We expect the classifier to \emph{learn} to discriminate between EEG epochs recorded while an irrelevant number is shown on the screen (negative class), and epochs recorded while the target number that is shown.

\begin{figure}[tb]
	\centering
	\includegraphics[width=0.7\columnwidth]
	{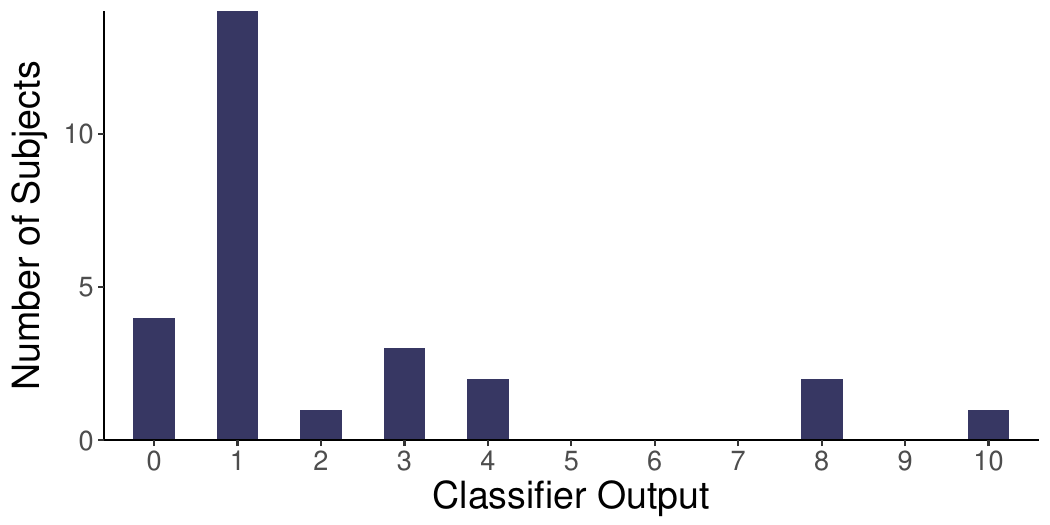}
	\caption{
		Aggregated output of a classifier trained on supraliminal data from the second counting experiment, conducted 20 minutes later.
		The classifier must extract the private information: which number between 0 and 10 was the user counting.
		While the achieved performance is not perfect, the classifier shows persistency over time, especially considering that part of classification errors are likely due to (measured) difference in participants' focus.
	}
	\label{fig_count_count2}
\end{figure}

\begin{figure*}[t]
	\captionsetup[subfigure]{justification=centering}
	\centering
	\hfill \null
	\subfloat[Aggregated outputs.] {
		\includegraphics[width=0.7\columnwidth]
		{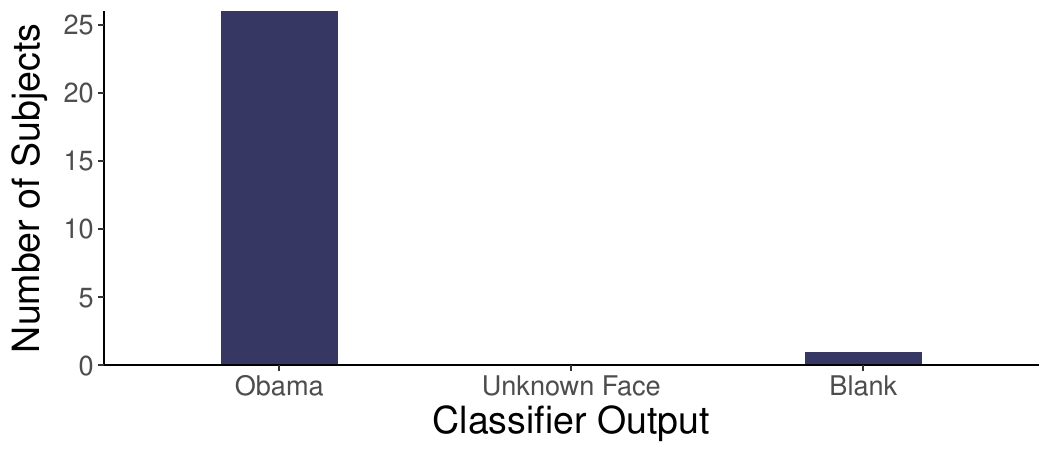}
		\label{fig_face_face}
	}
	\hfill
	\subfloat[Outputs depending on the level of stimulus detection.] {
\includegraphics[width=0.7\columnwidth]{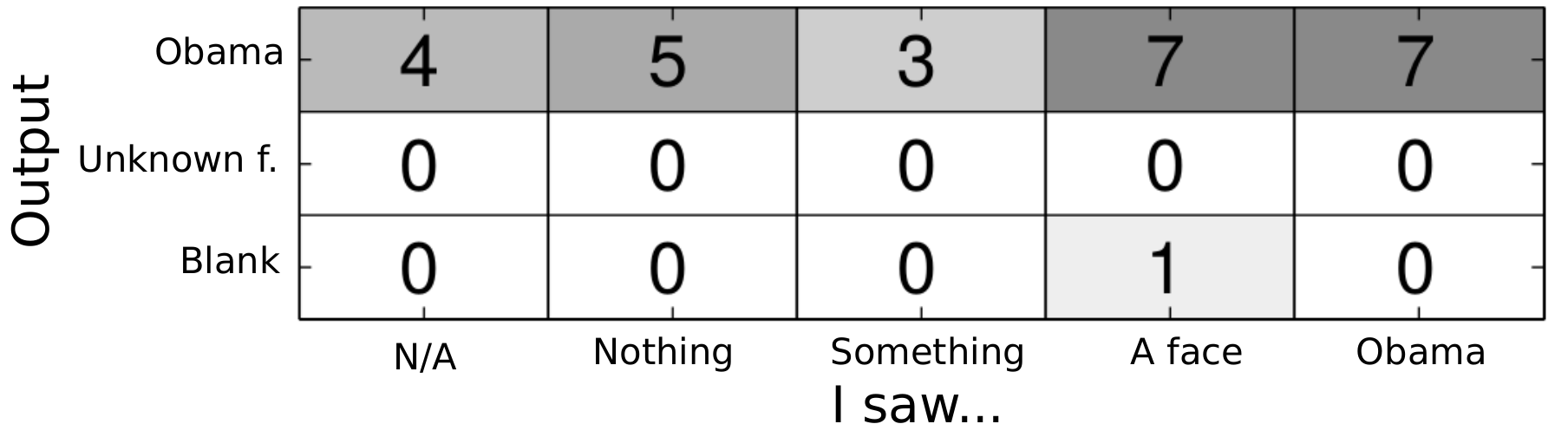}
		\label{fig_face_ob}
	}
	\hfill \null
	\caption{
		Outputs of the classifier trained on the first half of subliminal stimuli in the form of faces hidden in the video, and tested on the other half.
		After training on subliminal data, the classifier outputs the correct result for the large majority of users, irrespective of the level of their stimulus detection.
		This shows that subliminal stimulation does indeed impact the recorded EEG data.
	}
	\label{expFaceTrain}
\end{figure*}

When testing its performance, our goal is to verify if the classifier correctly determines user's private information, namely the number that they were counting.
While all users in our experiment were instructed to count the occurrences of number 1, this information is not explicitly available to classifiers that are trained for each user.
We test the performance of the classifier by feeding it with epochs of EEG data recorded while one of stimuli (numbers) were shown on the screen.
For each input epoch, the classifier outputs a score that represents the likelihood of the epoch having been recorded while the number shown on the screen was indeed the one that user counted (positive class).
In order to make a final decision on the \emph{private} counted number, we average the output scores across each of the 11 different stimuli (0-10), and pick the number with highest average score as the final output.

The classifier is successful if it indeed generalizes on the epochs of the second counting sequence, in which the user repeated the counting task.
The only major difference in the second counting sequence is that the users are slightly less focused after having conducted the calibration phase and after having watched the video.
This is supported by the average counting error of the users, which are larger in the second counting phase (0.72) than in the first phase (0.38).

\begin{figure}[t]
	\centering
	\includegraphics[width=0.75\columnwidth]
	{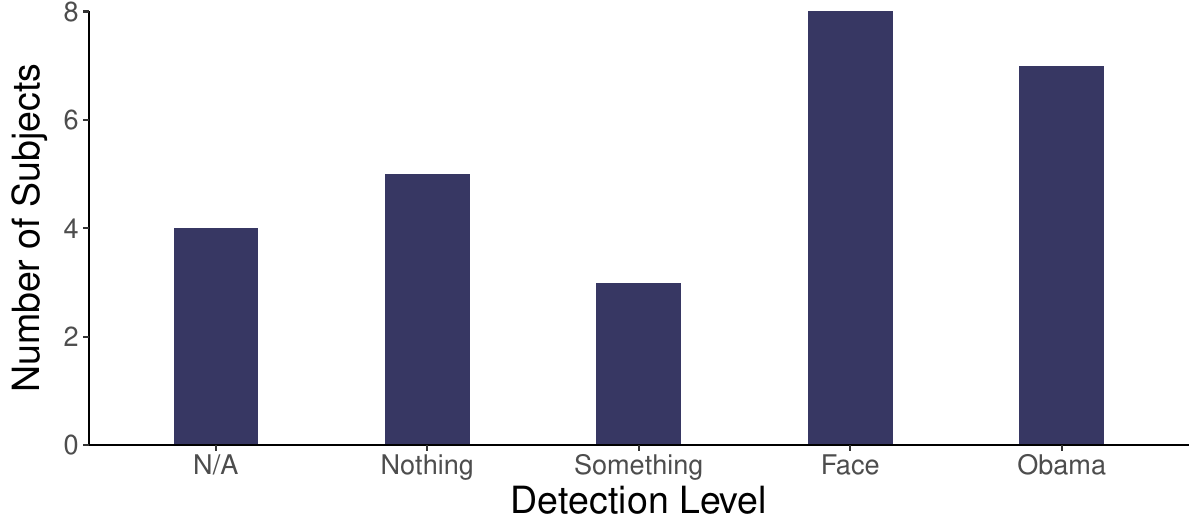}
	\caption{
		Aggregated levels of participants' detection of the visual stimuli hidden in the video. While the stimulus did not remain hidden from all participants, the stimulation was indeed subliminal, with only 7 participants able to correctly detect the stimuli that was repeatedly shown throughout the duration of the 20 minute long video.
	} \label{fig:DetectionLevels}
\end{figure}

\myparagraph{Results}
As a sanity check, we test the classifier both on data from the first counting phase, used for training, and the test data from the second counting phase.
As expected, the outputs on the training data are accurate: the classifier correctly guesses the private information (number 1) for all users.
Figure~\ref{fig_count_count2} shows the results of testing the classifier on the second counting sequence, recorded 20 minutes later.
Even though the accuracy of classification does decrease, \emph{the extracted private information is still correct for the majority of users}.
We conclude that the classifier indeed probabilistically extracts private information, but observe that detecting the relevant stimulus for a user in such counting experiment is not straightforward, even when the classifier is both trained and tested on supraliminal data.

In order to allow comparisons with related work on the topic of EEG-based privacy attacks~\cite{Usenix12_EEGattacks}, we quantify the attacker's information gain in each experiment as the relative reduction in Shannon entropy that the classifier output provides in comparison to the entropy of a random guessing attack.
The results show that, in comparison to randomly guessing the relevant stimuli, the attacker who relies on such classifier is able to \emph{reduce his guessing entropy by about 38.8\%}.
This is very much in line with previously reported EEG-based attacks on privacy~\cite{Usenix12_EEGattacks} that relied on supraliminal stimuli and allowed the attacker to achieve a 10-40\% decrease in guessing entropy.
Furthermore, these results suggest that, on a technical level, predicting relevant stimuli is a difficult learning task, such that any success that we achieve in our proof-of concept subliminal attack is significant.

\subsection{Stimulus Subliminality} \label{ssec:stimulusSubliminality}
We begin exploring subliminal stimulation by first analyzing the level to which the stimulus remained hidden from participants in the experiment.
As described in Section~\ref{sec:neuroBackground}, the level of cognitive perception of a visual stimulus depends on a number of factors which vary significantly between scenarios, environments, and individuals.
As a result, there usually do not exist established duration thresholds that result in subliminal responses.
Instead, subliminal stimuli are defined as those that remain undetected in a certain percentage of the times that they are shown to an individual. 
Given the fact that the same stimulus is repeatedly shown to participants in our experiment, we expect some of them to be able to detect it, even if we shorten the stimulus duration in comparison to what is often reported in relevant literature (10~ms to 55~ms).

The results of asking participants if anything seemed ``strange'' while watching the video are given in Figure~\ref{fig:DetectionLevels}.
The subjects are divided into five different groups, representing the different recognition levels of the users.
As expected, the subliminality of the stimulus varied for different users.
While a total of 7 participants were indeed able to recognize an image of Barack Obama, 5 users did not notice anything unusual, while further 3 of them only ``saw something''.
Despite being detected by a subset of users, our stimuli remained completely hidden from some users.
This supports the hypothesis that private information could be subliminally probed using EEG based BCI devices, especially if the attacker can adapt to a specific victim by gradually increasing the stimulus duration in order to ensure that it remains undetected.
We further discuss possible ways to hide the stimuli in Section~\ref{sec_discuss}.

\subsection{Probing for Subliminal Information}\label{ssec:probingSubliminalInfo}

Ultimately, the attacker's goal is to be able to train the classifier on EEG data that was recorded while the user was calibrating the device, for instance, by counting occurrence of a specific number (supraliminal stimuli).
Such classifier could then ideally be used on subliminal data to probe for relevancy of particular stimuli that are shown beyond victim's cognitive perception (e.g. to detect recognition of a face or detect which face is most relevant to the victim).
Before evaluating the classifier in this scenario, we first evaluate a variant of attack in which the adversary trains the classifier on known subliminal stimuli, and then probes the user using the same type of subliminal stimulus.
While training on supraliminal data is more likely to generalize well and be available since it is can be performed during standard calibration processes, training and both testing on the same domain and using subliminal data is useful to an attacker since it can offer a good proxy measure of confidence in classifier output.

\myparagraph{Setup}
We carry out an experiment in which we train the classifier on the first half of the epochs that have been recorded while the user watches the modified video.
This time, we label all epochs during which an Obama image was shown as the positive class, and epochs that correspond to blank video or unknown face as the negative class.
We call this modified classifier BLR$_F$ to account for the fact that it was trained on faces instead of numbers.
Hence, the classifier is directly trained for a specific type of stimuli and is, as such, expected to generalize to novel situations less successfully.

We test the classifier on the second half of the epochs collected from the modified video, using the same procedure as described in Section~\ref{ssec:classifierValidation}, where outputs for each of three stimuli classes are averaged for each user, and the maximum is taken as the final verdict.

\myparagraph{Results}
Figure~\ref{fig_face_face} shows the results of training on subliminally shown stimulus, and testing on subsequent epochs recorded while the same type of stimulus is presented.
The attack was successful for almost all users, failing only for one user who recognized a face in the video.
The statistics of the aggregated users across different levels of consciousness are shown in Figure~\ref{fig_face_ob}, where it is visible that achieved results do not vary significantly depending on the level of observation reported by users.
This on-the-fly calibration of the attack leads to very accurate results and has the advantage that the user is not required to perform a calibration phase.

The shown results allow us to conclude that subliminal stimuli do indeed trigger responses that are consistently detectable using EEG-based BCI devices.
Consequently, if the adversary is able to train a classifier while the user is exposed to similar stimuli for which private information is known, subliminal stimuli can pose a privacy threat to users.

\begin{figure}[t]
	\centering
	\includegraphics[width=0.7\columnwidth]
	{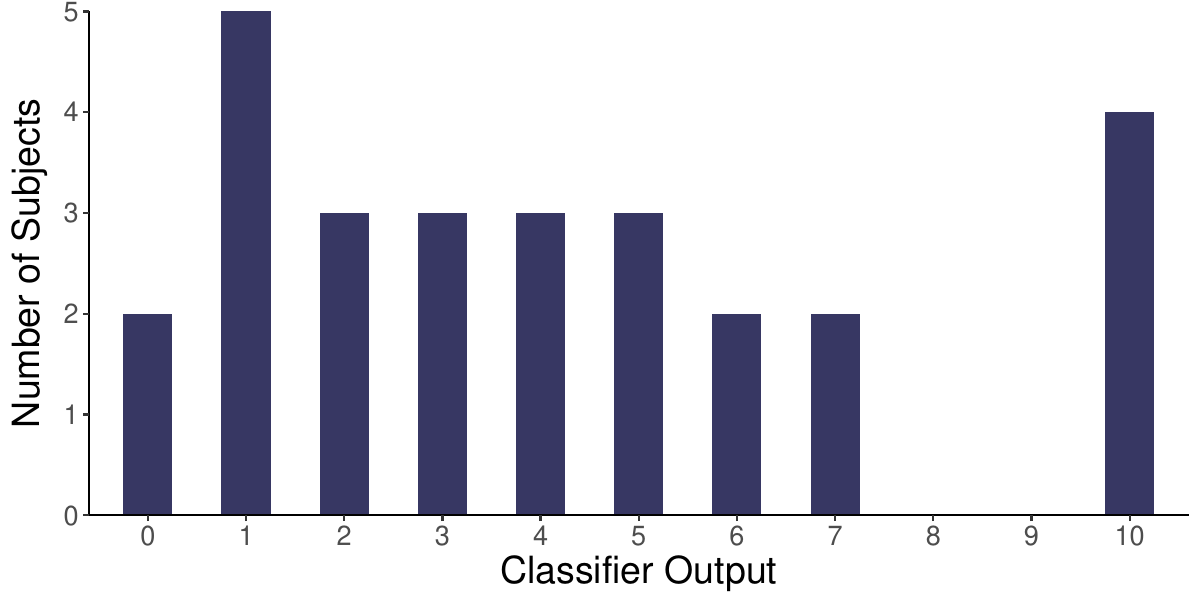}
\caption{
		Aggregated outputs of the classifier trained on subliminal faces stimuli, and tested on the counting sequence, while users counted occurrences of number 1.
		The classifier trained on subliminal data does not generalize on the supraliminal counting sequence, suggesting overlearning.
	}
	\label{fig_face_count1}
\end{figure}

\myparagraph{Generalization Performance}
We now evaluate how well does training on specific stimuli for subliminal face recognition generalize to other tasks, expecting generalization to suffer due to overlearning.
We investigate this hypothesis by testing the face-trained classifier BLR$_F$ on the epochs of the counting sequence and report the results in Figure~\ref{fig_face_count1}.
Even though the BLR$_F$ classifier has the biggest chance to output the correct answer (number 1), it also predicts  other numbers with similar probability.
It is visible that BLR$_F$ does not generalize well from subliminal training to testing on epochs recorded in counting experiments.

This means that the face-trained classifier BLR$_F$ might be less applicable than the classifier based classifier BLR.
A face in the visual field triggers many face-detection and face-recognition processes~\cite{ffa} and, as a consequence, leads to strong event-related potentials that might be detected by the BCI.
The weak generalization from face-based subliminal training to counting data is likely a result of BLR$_F$ classifier predominately using signals related to processing images of faces, which do not generalize well to other tasks.

\begin{figure*}[t]
  \captionsetup[subfigure]{justification=centering}
  \centering
  \hfill \null
  \subfloat[Aggregated outputs (Counting on Faces).] {
    \includegraphics[width=0.7\columnwidth]
                    {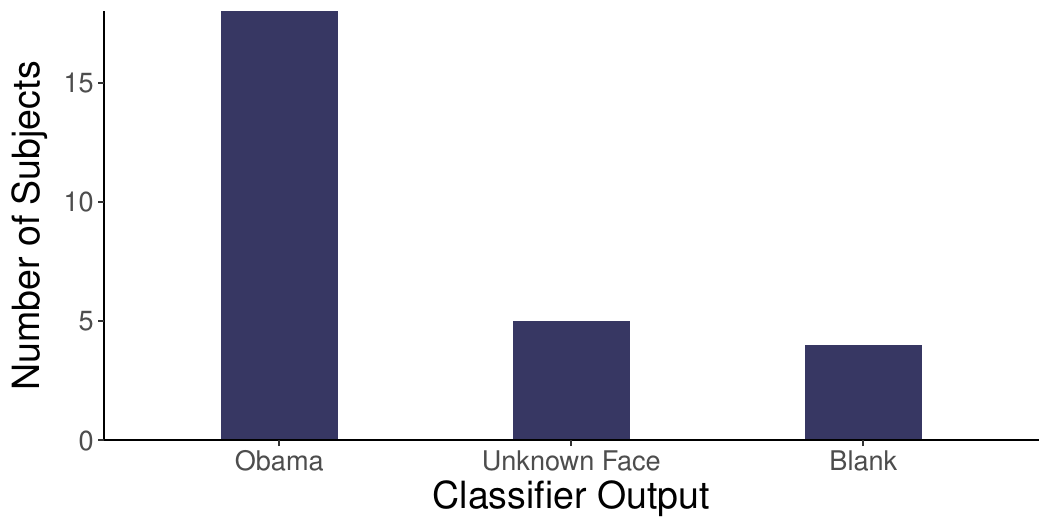}
    \label{fig_count_face_stats}
  }
  \hfill
  \subfloat[Outputs depending on the recognition level of the user.] {
    \includegraphics[width=0.7\columnwidth]{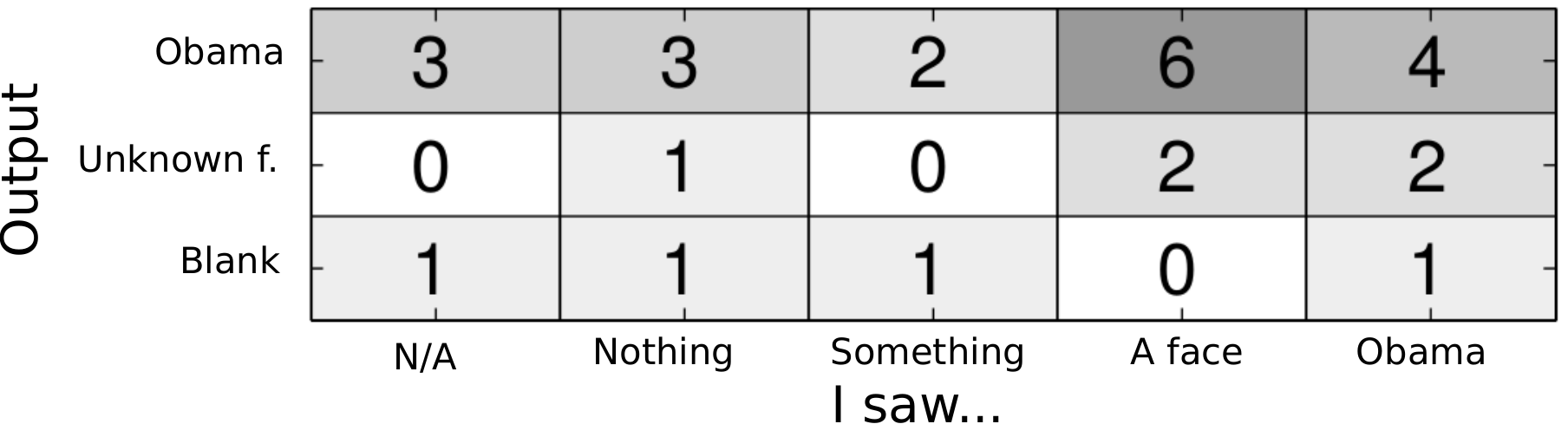}
\label{fig_count_face}
  }
  \hfill \null 
  \caption{
    Detecting subliminal responses based on supraliminal training (Numbers on Faces).
    For each user, the classifier outputs which type of stimulus results in recognition, the three possible candidates being: Obama, Unknown face, and random periods with no subliminal stimuli (Blank).
    Outputs are correct for the majority (18 out of 27) of participants (Figure a), irrespective of their level of stimuli detection (Figure b), showing that such an attack is indeed feasible.
    }
  \label{whole_fig}
\end{figure*}

\subsection{Transfer Learning: Subliminal Probing Using Supraliminal Training} \label{ssec:transferLearning}
After confirming that the subliminal responses can indeed be reliably detected using our EEG setup, we now evaluate the attack that subliminally probes for user's private information.
The adversary trains the classifier on supraliminal data acquired during user's calibration, and then uses the classifier on the subliminal data acquired while the user is performing a different task.
From a machine learning perspective, this is a challenging task, as the classifier needs to generalize from one domain to another.

\myparagraph{Setup}
During training, the BLR classifier was provided with data from Counting I, as was described in Section~\ref{ssec:classifierValidation}.
For testing, we extract all epochs triggered by hidden images of Barack Obama, all epochs triggered by the unknown face, and equally many epochs taken from random frames where the video was not manipulated.
Again, we let the classifier output a score for each epoch of this dataset.
Recall that, based on the training data used, this score outputs the classifier's belief that the user has `counted' the respective stimulus.
Even though the user did not actively count the target stimulus (she should not even realize that it is on the screen), the classifier is searching for the same artifacts in the EEG signal.

As in the counting experiment, the final output of the classifier is the candidate stimulus that gets the highest average classifier score.
This time, there are three possible outcomes.
Since we assume that all participants recognize an image of Barack Obama, we can compare the classifier output against this ground truth.

\myparagraph{Results}
We show the output statistics in Figure~\ref{fig_count_face_stats}.
For 18 users the classifier outputs the correct answer.
For 5 users, BLR predicted `Unknown face' and for 4 users BLR predicted `Blank'.
From a machine learning perspective, it appears that the attack works, as the classifier is able to distinguish a relevant stimulus from irrelevant stimuli.
The \emph{reduction in guessing entropy} is expectedly smaller than in previously reported supraliminal attacks (and our baseline); however, it still \emph{equals a high 20.84\%.}
This is an important result, which shows that attackers could indeed carefully design their visual stimuli such that they remain subliminal, and still probabilistically reduce the entropy of guessing relevant private information using EEG-based BCI devices.

We show the classifier output split by different levels of user awareness in Figure~\ref{fig_count_face}.
The attack works almost independently of the extent to which the victims realize that the video has been manipulated and in each recognition group, the classifier found the correct answer for the majority of users.
In this light, it is interesting to compare the results of BLR with the results of BRL$_F$, trained on face data and tested on counting data.

\myparagraph{Comparison of BLR and BLR$_F$}
The fact that the predictive power of the classifier translates from the counting scenario to the attack scenario, but not vice versa, suggests that the two classifier variants are trained to detect different neuro-physiological processes.
On the one hand, considering that BLR$_F$ classifier does not generalize well on the counting task, it is possible that the majority of its classification success stems from using face detection features, rather than detecting the stimulus that is most relevant to the user.
On the other hand, even though the BLR classifier is never trained on subliminal stimuli of recognizing faces (and hasn't been exposed to face recognition during training), it still achieves strong generalization performance.
This further gives confidence that the standard calibration task in which users are required to count one number suits as a strong basis for training classifiers that extract subliminal private information based on its relevancy to the victim.

\begin{figure}[t]
  \centering
  \includegraphics[width=0.7\columnwidth]
                    {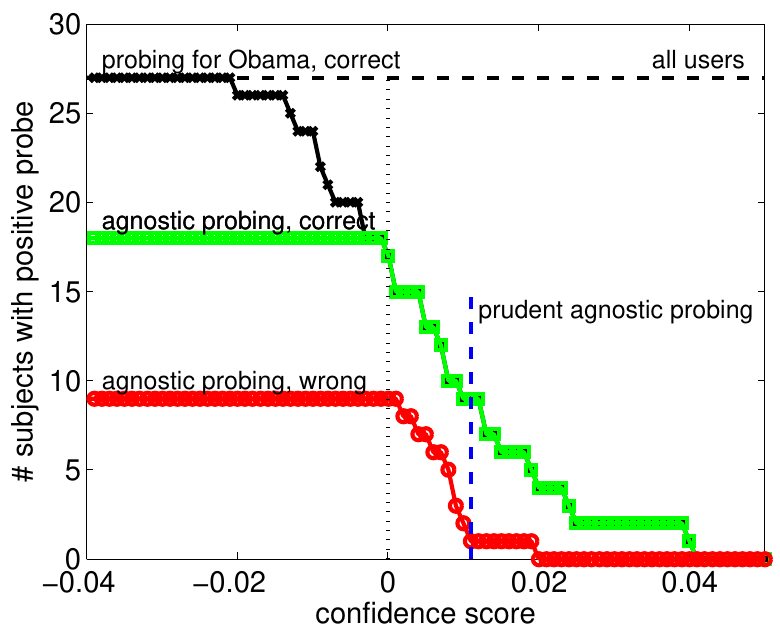}
    \caption{
    Using confidence thresholds, the attacker is able to increase classification performance by not making some decisions.
    Green and red lines depict correct and incorrect classifications depending on different thresholds.
    Black line shows the number of positively confirmed hypothesis that the most prominent stimulus is the one for known face.
    The vertical dashed line represents the threshold used in further analysis.
  } \label{fig_confPred}
\end{figure}

\begin{figure*}[t]
  \captionsetup[subfigure]{justification=centering}
  \centering
  \null \hfill
  \subfloat[Aggregated outputs.]{
    \includegraphics[width=0.7\columnwidth]
                    {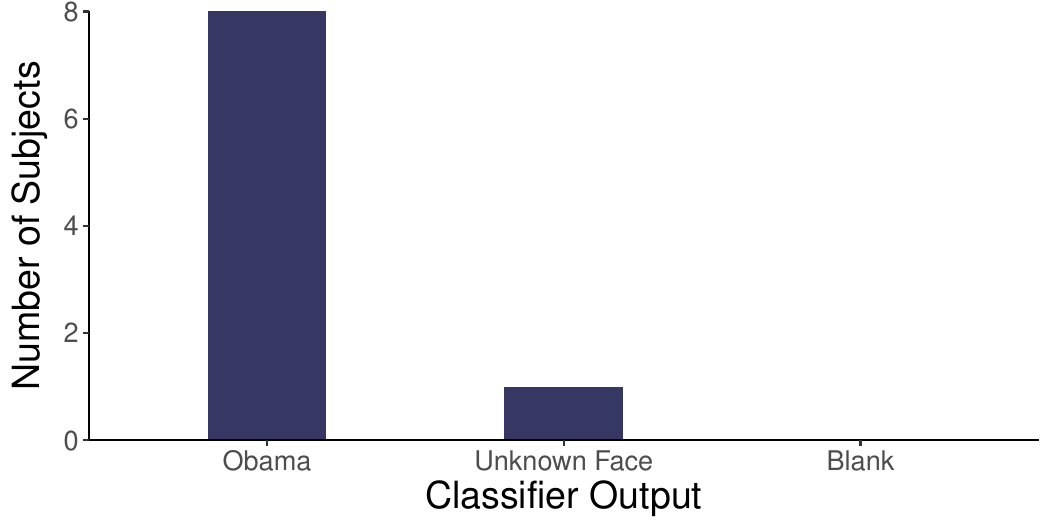}
    \label{fig_confStat}
  }
  \hfill
  \subfloat[Outputs depending on the stimuli detection levels.]{
    \includegraphics[width=0.7\columnwidth]
                    {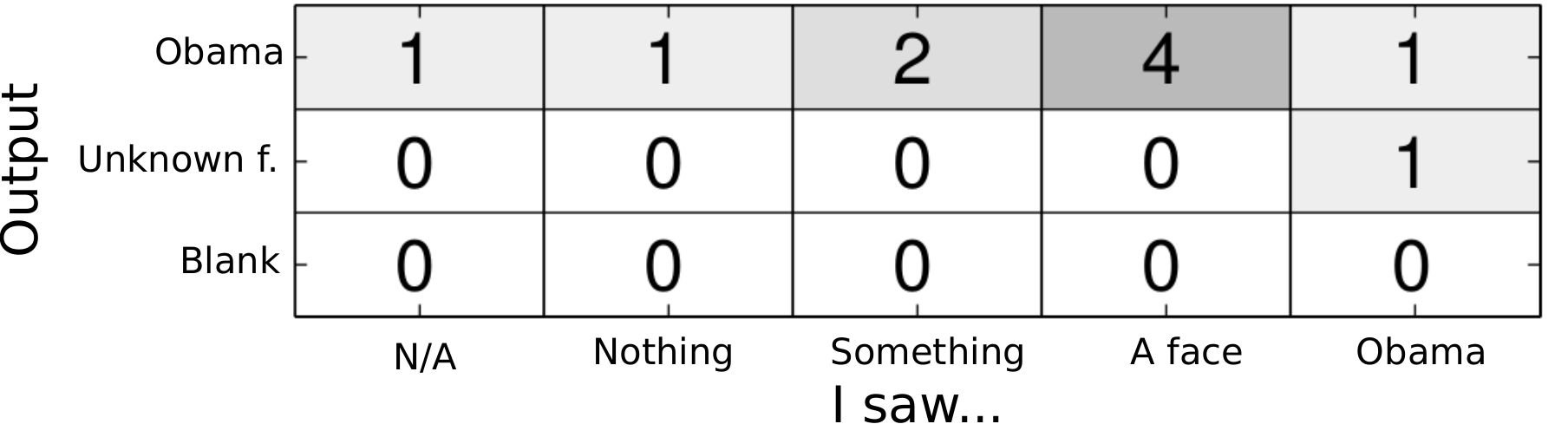}
    \label{fig_confDetail}
  }
  \hfill \null 
  \caption{Using stricter confidence thresholds, the attacker can trade making less decisions for increasing overall classification performance.
  For example, using a confidence threshold defined in Figure~\ref{fig_confPred} results in making correct classification decisions for 8 out of a total of 9 participants.
  }
  \label{fig:overallConfidenceResults}
\end{figure*}

\subsection{Utilizing Classification Confidence} \label{ssec:classificationConfidence}

We now propose a method to make the results of the attack more reliable, given the assumption that the attacker has an estimate of confidence in classifier outputs acquired by validation.
We begin by distinguishing two slightly different attack scenarios and then introduce a way to compute confidence scores for each of them.

\myparagraph{Targeted probing versus agnostic probing}
The experiments with the modified video are targeted towards a specific person that the attacker suspects the user might recognize and probes to validate this hypothesis.
He contrasts EEG data showing images of this person against images that are most likely irrelevant for the user.
If the classifier ranks the epochs highest that correspond to the person of interest, then the attacker concludes that the user knows this person.
In an alternative attack scenario, the attacker does not know for which alternative to test for.
In this \emph{agnostic} attack, the attacker confronts the user with $k$ stimuli that could potentially be relevant to the user.
Then the attacker must make a one-out-of-$k$ decision based on the recorded EEG data.
This task is more difficult than the targeted attack.
In the targeted attack the attacker can simply discard the hypothesis if the target stimulus did not achieve the highest score.
In agnostic probing, there will always be a stimulus that achieves the highest score.
The attacker must decide if he trusts this outcome.
In the next paragraph we propose a method to  decide with confidence, and report on the results of an attack where we simulate that the attacker does not know in advance that he is probing for Obama.

\myparagraph{Improving attacks using confidence measures}
In both the targeted attack and the agnostic attack, the attacker must decide whether to trust the classifier outcome and accept the hypothesis or to reject it.
A technical way to base this decision is to compute a measure of confidence for a given classifier outcome and request a minimal confidence score for accepting the hypothesis.

For targeted attacks, we propose to compute the difference between the average classifier scores of the epochs that show the target class and the average score of the closest non-target class.
For instance, if we probe for Obama's image and this class achieves the highest score, while the unknown face achieves the second best score, then the measure of confidence is the difference between these two scores.
Only if this difference exceeds a predefined threshold does the attacker accept it.
In all results that we have reported so far, we used a threshold of $0$, i.e. the Obama hypothesis is accepted if only the corresponding epochs score the highest.
In a scenario where the attacker wants to avoid false positives, he can implement a higher threshold to get more prudent estimates.
In the opposite situation, if the attacker is less risk-averse, he can use a negative confidence threshold to accept the hypothesis even if the target class is not ranked first by the classifier.

For agnostic attacks, the confidence score can be applied similarly.
If the class that ranks highest has a classifier score that exceeds the second best class by a predefined confidence score, then the attacker accepts the hypothesis. If not, the outcome means that none of the presented stimuli is relevant for the victim.

\myparagraph{Results}
In Figure~\ref{fig_confPred}, we report the results of these modified attacks.
The x-axis depicts the applied threshold of the confidence score.
The y-axis depicts the number of users for which the attack hypothesis was accepted by the attacker, given the current threshold.
With agnostic probing, incorrect hypotheses can also be accepted (we aggregate all incorrect outcomes, namely the unknown face and frames with no stimuli, under a single error rate).
If the confidence threshold is negative, the attacker accepts the Obama hypothesis even if its epochs do not achieve the highest average classifier score.

As can be seen, the number of subjects where the attacker accepts the correct hypothesis declines for all attacks as the threshold of the confidence score increases.
Luckily (for the attacker), the empirical chance of accepting a wrong answer declines considerably faster than the one of accepting correct answers.
As a result, there are configurations in which users can be attacked with better accuracy.
When an attacker deploys the attack, he must reason about risks and costs of false negatives and false positives and select an appropriate threshold.
For instance, in Figure~\ref{fig_confPred}, we highlight a threshold that would be a prudent choice for an agnostic attack where false positives are costly for the attacker.

The summary statistics of the outcome of such attacks with a prudent confidence threshold are given in Figure~\ref{fig_confStat}, while the detailed outcomes are depicted in Figure~\ref{fig_confDetail}.
As can be seen, the confidence criterion affects decisions at all levels of user awareness.
If the attacker is willing afford to not make a decision for all victims, he can in turn get more reliable results by apply the confidence criterion.
For instance, for a threshold of about 0.015, the attacker's probability of successful guessing increases to 8 out of 9 attempts, which is a \emph{decrease in entropy of 49.8\% in comparison to random guessing}.

\subsection{Summary of Experimental Results}

The main experimental results presented in this paper are threefold.

\textbf{Firstly}, in Section~\ref{ssec:stimulusSubliminality}, we show that a carefully designed visual stimulus can remain subliminal for a subset of users, while still eliciting detectable and consistent responses~(Section~\ref{ssec:probingSubliminalInfo}).

\textbf{Our main result} is presented in Section~\ref{ssec:transferLearning}.
Experiments show that a classifier that is trained on supraliminal data (recorded during user calibration) can indeed be useful to detect user's responses to subliminal stimulation.
This is an important result, since it shows that the attacker can reduce his guessing entropy on the task in our experiments by as much as 20\%.

\textbf{Finally}, our experiments use a conservative setup, in which both the classification threshold and the stimulus duration are identical for all users.
As results in Section~\ref{ssec:classificationConfidence} show and we further discuss in Section~\ref{sec_discuss}, the attacker can additionally increase his attack success by relying on measurements of classifier confidence and adapting stimulus duration to each victim.
For one such threshold, the classifier in our experiments makes the correct decision for 8 our 9 users, which effectively reduces attacker's guessing entropy by almost 49\%.

\section{Discussion} \label{sec_discuss}
We now critically discuss our attack and its possible extensions, and suggest several potential ways to defend against subliminal probing for private information.

\subsection{Extensions \& Limitations}

\myparagraph{Long-term attacks}
Although our experiment has shown that it is feasible within one EEG recording session to acquire information about the subject, the ideal setup in a real-world situation would be to collect EEG data over long periods of time.

The subliminal nature of the stimuli presents a large advantage over the method presented in Martinovic et al. in that the user may never realize anything strange is going on, and proceed to use the application and expose sensitive EEG data for large periods of time~\cite{Usenix12_EEGattacks}.  A larger body of data could be more useful in training the classifier and gaining more information about the user by making it possible to present a larger amount of subliminal stimuli.  It would therefore be easier to slowly build a profile of various facts about the user which could be used against them.

However, other EEG security work on using EEG as a form of authentication shows that EEG over multiple sessions may decrease in accuracy, as electrode placement and small aspects of the device setup can change from session to session~\cite{authentication}.
To address this problem, our method of subliminal validation with stimuli that the user certainly knows can be used to estimate the degradation of the classifier performance and trigger its re-training if needed.
\vspace{1em}

\myparagraph{Dry EEG}
Our current experiment uses research-grade EEG recording equipment, which is relatively expensive and has a long setup process involving the injection of gel into the EEG cap.
Additional experiments need to be done in order to conclude that similar effects can be captured with current consumer-grade EEG devices, but given the fact that we use only a small subset of channels that are found on consumer-grade devices as well, it is reasonable to assume that such attacks could be replicated.
Companies are looking to create EEG devices with dry electrodes that have the same resolution as ones that require gel~\cite{mynd}, since this reduces the setup time and allows performing more extensive experiments.
As such devices are made affordable to the public, it becomes more likely that subliminal attacks similar to ones presented in this paper are deployed.

\myparagraph{Improved strategies to hide stimuli}
One direction in which our proof-of-concept could be improved is to better hide the stimuli in the video that is being shown.
With respect to subliminal stimulation, we are constrained by the need to choose the same duration for each individual, as well as the limits of the hardware used by the victim.
As already discussed, the attacker could adapt the stimuli duration for each specific user and thus ensure that it remains subliminal.

Furthermore, the subliminal effect can be increased through non-temporal means in several ways.
For instance, stimuli that are presented in the foveal point (the position of user's fixation) require significantly shorter stimulus presentation in order to remain subliminal then parafoveal stimuli, which are approximately one to five degrees from fixation.
Such stimulation can maintain subliminal effects with a much longer stimulus time~\cite{subliminal_review,foveal} and could be utilized if the attacker knows where the victim's gaze is focused, for instance by using a gaze tracking device.

Another factor affecting the subliminal effect is the continuation of the visual processing of a stimulus, long after the stimulus has disappeared from the user's sight.
After a stimulus is shown on the screen, the image can still remain on the user's retina for duration of up to 30 milliseconds~\cite{retina}.
During this additional time the visual information is still being sent to the brain. This means the retina serves as a buffer that undermines the attackers effort of hiding the stimulus. Particularly, if the face image is shown on a low entropy background that stays calm for an extended period of time, the time that the user can "see" the face exceeds the actual duration at which it was displayed on the screen.

It has been shown that by placing an even stronger stimulus directly after the respective stimulus one can overwrite this visual buffer, a technique named backward masking~\cite{masking}. Several forms of backward masking have been explored, ranging from bright flashes, patterns, and even noises, which would not block low-level visual processing, but could interrupt higher level vision processing in the brain~\cite{subliminal_review,masking}. In order to not raise the user's alertness, one cannot simply use another artificial image as a mask, of course. The art of masking a subliminal attack here would rather be to identify frames of the video/game that provide original sudden local changes in contrast, color, or sound. The stimulus could then be deployed directly before such an event occurs.

\subsection{Countermeasures}
We now discuss several potential countermeasures to attacks presented in this paper.

\myparagraph{Detecting rapid screen changes}
In order for the presented subliminal stimulation to be successful, the attacker needs to repeat the presentation of the stimulus multiple times, each time showing it for a very short duration.
Assuming that the EEG-based BCI device controls screen output, such suspiciously short changes could be detected and result in a warning shown to the user.
A similar example are web browsers that detect attempts of device fingerprinting using HTML5 Canvas and ask the user if such behavior should be allowed.

\myparagraph{Limiting access to raw data}
An obvious potential countermeasure is to limit or restrict access to raw EEG data provided by BCI devices.
In this scenario, the core BCI platform analyzes raw data and detects specific events, while different applications only receive input about user's behavior through a specific API.
While such architecture changes can significantly reduce potential privacy threats, they also reduce the ability of application developers to innovate in positive ways beyond what the BCI vendors implement as the interface.
Finally, adding noise to raw data could reduce the potential to extract weaker signals, such as those resulting from subliminal stimuli, but is likely to also reduce the performance of other third-party applications that rely on raw data.

\myparagraph{User awareness}
Recent research on the impact of subliminal advertising has shown that individuals who are warned of the presence of subliminal ads have significantly decreased effects of advertising compared to the control group~\cite{Verwijmeren2013}.
Furthermore, subliminal effects were reduced even in the case where the warning was given after the stimulation.
While no research has yet been done on the impact of such warnings on brain activation, these findings suggest that users could potentially be shielded against subliminal probing for private information if they were warned or aware that such possibility exists before they start using EEG-based BCI devices.

We believe that this paper plays an important first step in the role of raising awareness of the possibility of subliminal attacks among both the future users of EEG-based BCI devices, as well as device vendors, who are in position to pro-actively implement countermeasures to such threats.

\section{Related Work}
\label{sec_related}
In this section, we overview existing literature on neuroscientific aspects of security-relevant applications of EEG.

\myparagraph{Recognizing faces for authentication}
Any information about faces familiar to a user should be considered vulnerable private information as it is already being used for security purposes.
The human ability to recognize and remember faces over extended periods of time has been utilized as a method of authentication in which the user identifies familiar faces within a grid of images~\cite{passfaces}.
This method was shown to relieve users of the need to memorize word-based passwords, and instead rely on the natural ability to remember faces, even along a timespan of several months.
The idea of Passfaces was further extended by the use of commodity BCI devices~\cite{dunphy2008gaze}.
Instead of manually choosing the correct faces out of a grid of images, the authors used eye trackers and considered a 0.5 second fixation on a face as a selection of it.
Other applications of facial recognition are used in real-world situations, for instance security verification for Facebook users, which requires one to identify their friends in tagged photos~\cite{FBauth}.
On a similar note of using subliminal knowledge in a security context, Bojinov et al. proposed using \emph{implicit learning} to implement coercion-resistant passwords for user authentication~\cite{Bojinov:2012:NMC:2362793.2362826}.
As our experiments demonstrate, facial recognition can be detected through subliminal presentation of the stimuli, which poses a threat to such any such security features that rely on users identifying familiar faces.

\myparagraph{Subliminal face recognition}
Given that facial recognition is has been repeatedly proposed as a method of authentication, it becomes even more important to test the possibilities of extracting information from facial stimuli.
Existing neuroscientific work on the subliminal perception of human faces shows that ERPs in response to unpleasant facial expressions have a higher positive amplitude than pleasant expressions.
Furthermore, this effect shows even through very fast unmasked subliminal presentations of stimuli, at 1ms~\cite{Bernat200111}.
This shows that visual information regarding faces can be processed and produce variance in the EEG signal even at a very subliminal level.
Although in our experiment several subjects had noticed the stimuli, a presentation time as little as 1~ms could potentially still reveal enough information in their EEG signal to extract desired information about faces~\cite{fear}.

\section{Conclusion} \label{sec_conclusion}
In this work we examined the feasibility of subliminal attacks on users of EEG-based brain-computer interfaces (BCIs).

In a series of experiments with 27 subjects, we find that our attack is able to detect brain responses to subliminal stimulation with accuracy that is comparable to the results previously reported for supraliminal attacks, even when the classifier is trained on a different type of brain responses than the ones that are being probed for.
As a consequence, by carefully designing the visual stimuli, an attacker can reduce the entropy of guessing user's private information by more than 20\%, while at the same time achieving that the victim remains unaware of being probed.

As a first attempt to perform subliminal probing, our experiments have been carried out in a controlled setting to demonstrate their feasibility and exclude other factors that might impede success.
Taking this into account, this paper also discusses attack improvements that can be achieved outside of a controlled lab environment and several potential countermeasures that users and manufacturers of EEG-based BCI devices can take to prevent such attacks in the future.

Given the recent improvements of measurement performance and the reduction of prices, the pervasiveness of EEG-based BCI devices in our daily lives is likely to increase.
Consequently, by experimentally showing the feasibility of such privacy compromises, we believe that this paper makes an important and timely contribution towards reducing the impending threats of attacks that include subliminal probing for private user information.

\bibliographystyle{IEEEtran}
\bibliography{bibfile}

\vskip 10pt
\begin{IEEEbiographynophoto}{Mario Frank}
	is a postdoc researcher at the Department of Electrical Engineering and Computer Science, University of California, Berkeley.
\end{IEEEbiographynophoto}

\begin{IEEEbiographynophoto}{Tiffany~Hwu}
	is a graduate student at the School of Social Sciences, University of California, Irvine.
\end{IEEEbiographynophoto}

\begin{IEEEbiographynophoto}{Sakshi~Jain}
	is a senior data scientist at LinkedIn Corporation.
\end{IEEEbiographynophoto}

\begin{IEEEbiographynophoto}{Robert~T.~Knight, M.D.}
	is a Professor of Psychology and Neuroscience  at the Department of Psychology, University of California, Berkeley.	
\end{IEEEbiographynophoto}

\begin{IEEEbiographynophoto}{Ivan~Martinovic}
	is an Associate Professor at the Department of Computer Science, University of Oxford.
\end{IEEEbiographynophoto}

\begin{IEEEbiographynophoto}{Prateek~Mittal}
	is an Assistant Professor at the Department of Department of Electrical Engineering, Princeton University.
\end{IEEEbiographynophoto}

\begin{IEEEbiographynophoto}{Daniele~Perito}
	is a postdoc researcher at Department of Electrical Engineering and Computing, University of California, Berkeley.
\end{IEEEbiographynophoto}

\begin{IEEEbiographynophoto}{Ivo~Sluganovic}
	is a graduate student at the Department of Computer Science, University of Oxford.
\end{IEEEbiographynophoto}

\begin{IEEEbiographynophoto}{Dawn~Song}
	is a Professor at the Department of Computer Science, 
	University of California, Berkeley.
\end{IEEEbiographynophoto}

\end{document}